%

\magnification 1200

\hsize= 16 truecm
\vsize= 22 truecm

\hoffset=-0.0 truecm
\voffset= +1 truecm

\baselineskip=18 pt

\parindent 10pt 

\footline={\iftitlepage{\hfil}\else
       \hss\tenrm-- \folio\ --\hss\fi}	

\parskip 0pt 


\font\bello= cmr10 scaled \magstep3
\font\piccolo=cmr8

\font\pbf=cmbx8

\def\eq{\autoeqno}
\def\re{\eqrefp}


\def\eq#1{\autoeqno{#1}}
\def\re#1{\eqrefp{#1}}

\newcount\notenumber \notenumber=1
\def\nota#1{\unskip\footnote{$^{\the\notenumber}$}{\piccolo #1}%
  \global\advance\notenumber by 1}

\def\s{\scriptstyle}  

\catcode`@=11 
%

\newcount\cit@num\global\cit@num=0

\newwrite\file@bibliografia
\newif\if@bibliografia
\@bibliografiafalse

\def\lp@cite{[}
\def\rp@cite{]}
\def\trap@cite#1{\lp@cite #1\rp@cite}
\def\lp@bibl{[}
\def\rp@bibl{]}
\def\trap@bibl#1{\lp@bibl #1\rp@bibl}

\def\refe@renza#1{\if@bibliografia\immediate        
    \write\file@bibliografia{
    \string\item{\trap@bibl{\cref{#1}}}\string
    \bibl@ref{#1}\string\bibl@skip}\fi}

\def\ref@ridefinita#1{\if@bibliografia\immediate\write\file@bibliografia{ 
    \string\item{?? \trap@bibl{\cref{#1}}} ??? tentativo di ridefinire la 
      citazione #1 !!! \string\bibl@skip}\fi}

\def\bibl@ref#1{\se@indefinito{@ref@#1}\immediate
    \write16{ ??? biblitem #1 indefinito !!!}\expandafter\xdef
    \csname @ref@#1\endcsname{ ??}\fi\csname @ref@#1\endcsname}

\def\c@label#1{\global\advance\cit@num by 1\xdef            
   \la@citazione{\the\cit@num}\expandafter
   \xdef\csname @c@#1\endcsname{\la@citazione}}

\def\bibl@skip{\vskip 0truept}


\def\stileincite#1#2{\global\def\lp@cite{#1}\global
    \def\rp@cite{#2}}
\def\stileinbibl#1#2{\global\def\lp@bibl{#1}\global
    \def\rp@bibl{#2}}

\def\citpreset#1{\global\cit@num=#1
    \immediate\write16{ !!! cit-preset = #1 }    }

\def\autobibliografia{\global\@bibliografiatrue\immediate
    \write16{ !!! Genera il file \jobname.BIB}\immediate
    \openout\file@bibliografia=\jobname.bib}

\def\cref#1{\se@indefinito                  
   {@c@#1}\c@label{#1}\refe@renza{#1}\fi\csname @c@#1\endcsname}

\def\cite#1{\trap@cite{\cref{#1}}}                  
\def\ccite#1#2{\trap@cite{\cref{#1},\cref{#2}}}     
\def\cccite#1#2#3{\trap@cite{\cref{#1},\cref{#2},\cref{#3}}}
\def\ccccite#1#2#3#4{\trap@cite{\cref{#1},\cref{#2},\cref{#3},\cref{#4}}}
\def\ncite#1#2{\trap@cite{\cref{#1}--\cref{#2}}}    
\def\upcite#1{$^{\,\trap@cite{\cref{#1}}}$}               
\def\upccite#1#2{$^{\,\trap@cite{\cref{#1},\cref{#2}}}$}  
\def\upncite#1#2{$^{\,\trap@cite{\cref{#1}-\cref{#2}}}$}  

\def\clabel#1{\se@indefinito{@c@#1}\c@label           
    {#1}\refe@renza{#1}\else\c@label{#1}\ref@ridefinita{#1}\fi}

\def\biblskip#1{\def\bibl@skip{\vskip #1}}           

\def\insertbibliografia{\if@bibliografia             
    \immediate\write\file@bibliografia{ }
    \immediate\closeout\file@bibliografia
    \catcode`@=11\input\jobname.bib\catcode`@=12\fi}


\def\commento#1{\relax} 
\def\biblitem#1#2\par{\expandafter\xdef\csname @ref@#1\endcsname{#2}}



%
%
\def\b@lank{ }


\newif\iftitlepage      \titlepagetrue

\def\titoli#1{
         \xdef\prima@riga{#1}\voffset+20pt
        \headline={\ifnum\pageno=1
             {\hfil}\else\hfil{\piccolo \prima@riga}\hfil\fi}}

\def\duetitoli#1#2{
                    \voffset=+20pt
                    \headline={\iftitlepage{\hfil}\else
                              {\ifodd\pageno\hfil{\piccolo #2}\hfil
             \else\hfil{\piccolo #1}\hfil\fi}\fi} }

\def\la@sezionecorrente{0}

\catcode`@=12


\autobibliografia

%
\catcode`@=11 
%
%
\def\b@lank{ }

\newif\if@simboli
\newif\if@riferimenti
\newif\if@bozze

\newwrite\file@simboli
\def\simboli{
    \immediate\write16{ !!! Genera il file \jobname.SMB }
    \@simbolitrue\immediate\openout\file@simboli=\jobname.smb}

\def\bozze{\@bozzetrue}

\newcount\eq@num\global\eq@num=0
\newcount\sect@num\global\sect@num=0

\newif\if@ndoppia
\def\numerazionedoppia{\@ndoppiatrue\gdef\la@sezionecorrente{\the\sect@num}}

\def\se@indefinito#1{\expandafter\ifx\csname#1\endcsname\relax}
\def\spo@glia#1>{} 

\newif\if@primasezione
\@primasezionetrue

\def\s@ection#1\par{\immediate
    \write16{#1}\if@primasezione\global\@primasezionefalse\else\goodbreak
    \vskip\spaziosoprasez\fi\noindent
    {\bf#1}\nobreak\vskip\spaziosottosez\nobreak\noindent}
%

\def\sezpreset#1{\global\sect@num=#1
    \immediate\write16{ !!! sez-preset = #1 }   }

\def\spaziosoprasez{26pt plus5pt minus3pt}
\def\spaziosottosez{15pt}

\def\sref#1{\se@indefinito{@s@#1}\immediate\write16{ ??? \string\sref{#1}
    non definita !!!}
    \expandafter\xdef\csname @s@#1\endcsname{??}\fi\csname @s@#1\endcsname}

\def\autosez#1#2\par{
    \global\advance\sect@num by 1\if@ndoppia\global\eq@num=0\fi
    \xdef\la@sezionecorrente{\the\sect@num}
    \def\usa@getta{1}\se@indefinito{@s@#1}\def\usa@getta{2}\fi
    \expandafter\ifx\csname @s@#1\endcsname\la@sezionecorrente\def
    \usa@getta{2}\fi
    \ifodd\usa@getta\immediate\write16
      { ??? possibili riferimenti errati a \string\sref{#1} !!!}\fi
    \expandafter\xdef\csname @s@#1\endcsname{\la@sezionecorrente}
    \immediate\write16{\la@sezionecorrente. #2}
    \if@simboli
      \immediate\write\file@simboli{ }\immediate\write\file@simboli{ }
      \immediate\write\file@simboli{  Sezione 
                                  \la@sezionecorrente :   sref.   #1}
      \immediate\write\file@simboli{ } \fi
    \if@riferimenti
      \immediate\write\file@ausiliario{\string\expandafter\string\edef
      \string\csname\b@lank @s@#1\string\endcsname{\la@sezionecorrente}}\fi
    \goodbreak\vskip 48pt plus 60pt
    \noindent\if@bozze\llap{\it#1\quad }\fi
      {\bf\the\sect@num.\quad #2}\par\nobreak\vskip 15pt
    \nobreak\noindent}

\def\semiautosez#1#2\par{
    \gdef\la@sezionecorrente{#1}\if@ndoppia\global\eq@num=0\fi
    \if@simboli
      \immediate\write\file@simboli{ }\immediate\write\file@simboli{ }
      \immediate\write\file@simboli{  Sezione ** : sref.
          \expandafter\spo@glia\meaning\la@sezionecorrente}
      \immediate\write\file@simboli{ }\fi
    \s@ection#2\par}


\def\eqpreset#1{\global\eq@num=#1
     \immediate\write16{ !!! eq-preset = #1 }     }

\def\eqref#1{\se@indefinito{@eq@#1}
    \immediate\write16{ ??? \string\eqref{#1} non definita !!!}
    \expandafter\xdef\csname @eq@#1\endcsname{??}
    \fi\csname @eq@#1\endcsname}

\def\eqlabel#1{\global\advance\eq@num by 1
    \if@ndoppia\xdef\il@numero{\la@sezionecorrente.\the\eq@num}
       \else\xdef\il@numero{\the\eq@num}\fi
    \def\usa@getta{1}\se@indefinito{@eq@#1}\def\usa@getta{2}\fi
    \expandafter\ifx\csname @eq@#1\endcsname\il@numero\def\usa@getta{2}\fi
    \ifodd\usa@getta\immediate\write16
       { ??? possibili riferimenti errati a \string\eqref{#1} !!!}\fi
    \expandafter\xdef\csname @eq@#1\endcsname{\il@numero}
    \if@ndoppia
       \def\usa@getta{\expandafter\spo@glia\meaning
       \la@sezionecorrente.\the\eq@num}
       \else\def\usa@getta{\the\eq@num}\fi
    \if@simboli
       \immediate\write\file@simboli{  Equazione 
            \usa@getta :  eqref.   #1}\fi
    \if@riferimenti
       \immediate\write\file@ausiliario{\string\expandafter\string\edef
       \string\csname\b@lank @eq@#1\string\endcsname{\usa@getta}}\fi}

\def\autoreqno#1{\eqlabel{#1}\eqno(\csname @eq@#1\endcsname)
       \if@bozze\rlap{\it\quad #1}\fi}
\def\autoleqno#1{\eqlabel{#1}\leqno\if@bozze\llap{\it#1\quad}
       \fi(\csname @eq@#1\endcsname)}
\def\eqrefp#1{(\eqref{#1})}
\def\numeriadestra{\let\autoeqno=\autoreqno}
\def\numeriasinistra{\let\autoeqno=\autoleqno}
\numeriadestra

\catcode`@=12

\numerazionedoppia


\titlepagetrue

\vglue 1 truecm

\centerline {\bello    Barriers between metastable states             }
\centerline {\bello         in the p-spin spherical model                    }

\vskip 1 truecm

\centerline {Andrea Cavagna, Irene Giardina, Giorgio Parisi}

\vskip 1 truecm

\centerline {\it          Dipartimento di Fisica}
\centerline {\it  Universit\`a di Roma I, "La Sapienza"}
\centerline {\it   P.le A. Moro 5, 00185 Roma, Italy }
\centerline {\it              and                    } 
\centerline {\it INFN Sezione di Roma I, Roma, Italy.}
\vskip 0.5 truecm

\centerline {\sl cavagna@roma1.infn.it}
\centerline {\sl giardina@roma1.infn.it}
\centerline {\sl parisi@roma1.infn.it}

\vskip 0.5 truecm

\centerline{February 7, 1997}

\vskip 2 truecm

\centerline{\bf Abstract}

\vskip 0.5 truecm

{\piccolo \noindent 
In the context of the $\s p$-spin spherical model for generalized spin glasses,
we give an estimate of the free energy barriers separating an equilibrium state
from the metastable states close to it. }

\vfill\eject


\titlepagefalse

\autosez{intro} Introduction.
\par

The structure of the phase space in the $p$-spin spherical model below the 
dynamical transition temperature $T_d$ is characterized by the presence of an 
exponentially high number of metastable states \ccite{kpz}{crisatap}. 
If we consider corrections to mean field theory when the size $N$ 
of the system is finite but large, the dynamics is determined essentially 
by two factors: the mutual disposition of the states and the free energy 
barriers between them.

The {\it real replica method} shed some light into the structure of the
metastable states of this model \cccite{kpz}{franzparisi}{noi} 
and the same method can be used to estimate the barriers.
It is now known that, given an equilibrium state, there are
many metastable states of various energies at finite overlaps with it; it is
therefore interesting to evaluate the free energy barriers between an 
equilibrium state and the metastable states close to it. 

In \cite{noi} we introduced a three replica potential: the first replica is
located into an equilibrium state, while the second one is constrained to stay
into a metastable state at given overlap with the first one (for particular
values of the overlap this second state can be of equilibrium too);  
we then calculated the free energy of a third replica as a function of its
distances from the other two and found two non trivial minima of this free 
energy, corresponding to replica 3 in equilibrium into the state of replica 1 
or into the state of 
replica 2. In this context, it is reasonable to give an estimate of the barrier
between these two states, following the free energy profile of replica 3 when 
it moves from a minimum to the other. This is the purpose of the present 
letter.

\autosez{uno} The method.
\par

The $p$-spin spherical model is defined by the Hamiltonian
$$
H(\sigma)=  \sum_{i_1<i_2<\dots<i_p} J_{i_1\dots i_p} 
\sigma_{i_1}\dots \sigma_{i_p}    
\eq{model}
$$
where the $\sigma$ are real variables satisfying the spherical constraint
${1\over N}  \sum_i \sigma_i^2=  1 $ and the couplings $J_{i_1\dots i_p}$ 
are Gaussian variables with zero mean and variance $p!\over 2 N^{p-1}$ 
\ccccite{grome}{gard}{tirumma}{crisa1}. We consider the case $p=3$.

The organization of the states in this model is quite complex 
\ccite{kpz}{crisatap}. 
In the range of temperature $T_c<T<T_d$ 
($T_c$ is the static transition temperature) the equilibrium states coincide
with those TAP solutions \cite{tap} which optimize the balance between the 
free energy 
$f$ and the complexity $\Sigma(f)$, i.e. solutions which minimize the 
function $\phi=f-T\Sigma(f)$; all other TAP solutions correspond to metastable 
states. 

To inspect the structure of all these states, i.e. their mutual overlaps,
we defined in \cite{noi} a three replica potential $V_3(q_{12}|q_{13},q_{23})$,
having the following features: replica 1 is  an equilibrium 
configuration of the system, while replica 2 is a typical configuration
of a system forced to equilibrate 
at overlap $q_{12}$ with 1; in other words, replica 2 chooses the most 
convenient configuration compatibly with the imposed constraint; $V_3$ is then
the free energy of a third replica 3, constrained to have overlaps 
$(q_{13},q_{23})$ with the first two quenched replicas.
We studied $V_3$ in the plane $\pi \equiv (q_{13},q_{23})$, at fixed value of 
$q_{12}$: 
a minimum of the potential in this plane corresponds to replica 3 having found
a state into which it can thermalize. 

In the temperature range $T_c < T < T_d$, at given $q_{12}$, $V_3$ has two non 
trivial minima, from here on $M_1$ and $M_2$:
{\parindent=1 truecm
\item{$\bullet$}
In $M_1$ replica 3 is located near replica 1, that is in the same equilibrium 
state; indeed, in $M_1$ the free energy and the self overlap of replica 3
satisfy the relation of TAP equilibrium solutions. 
\item{$\bullet$}
On the other hand, $M_2$
corresponds to replica 3 near replica 2 and its free e\-ner\-gy and self overlap
satisfy the relation of TAP metastable solutions; this means that replica 3
has thermalized into a metastable state at distance $q_{12}$ from the 
equilibrium state of replica 1.
\par}

Varying $q_{12}$, $M_1$ always corresponds to the same equilibrium state,
while $M_2$ identifies with different metastable states, at various distances.

The important thing is then that, fixed $q_{12}$, we have the possibility of 
studying the free energy barrier between two well defined states, via the
analysis of the free energy contour around $M_1$ and $M_2$ given by $V_3$.
In particular, we are interested in the barrier that has to be crossed to go 
from the equilibrium state represented by $M_1$ into the metastable state
represented by $M_2$. Indeed, the computation of this barrier is useful for
a comparison with the dynamical situation at finite $N$, in which a 
configuration starts at time zero from an equilibrium state and 
after an exponentially large time jumps to a metastable state. 
 
The plane $\pi$ is the image of the phase space $\Gamma$ through the mapping 
which, fixed $q_{12}$, maps a configuration $\sigma$ of $\Gamma$ 
into its distances $(q_{13},q_{23})$ from configurations 1 and 2. In this way,
a path in $\Gamma$ is mapped into a single path in $\pi$, while, obviously, 
the opposite does not hold. Due to this, a path in the plane $\pi$ could
correspond to no continuous path in the phase space and for this reason we can
only give a lower bound for the free energy barrier between the two states.

The proposal is to find in $\pi$ the path linking $M_1$ to $M_2$ 
which minimizes the variation of $V_3$ and to consider this variation as a 
lower bound for the free energy barrier between the two corresponding states.
To this aim it is clearly important to consider not only the minima of $V_3$ 
but also its saddles; indeed, as can be seen from next figures, the problem of
finding the best path from $M_1$ to $M_2$ reduces to single out the chain of 
saddles which minimizes the variation of $V_3$; the estimate of the barrier is 
then given by 
$$
\Delta F \geq \ V_3(S_{max}) - V_3(M_1)
\eq{tromba}
$$
where $S_{max}$ is the highest saddle crossed along the path.

\autosez{rip} The results.
\par

Before analyzing the structure of minima and saddles of $V_3$ at various values
of $q_{12}$, we remind that the minimum $M_2$ exists only in the range 
$0 \leq q_{12} \leq \bar q(T)$, and that for $q_{12}=q^\star (T)$, the state 
associated to $M_2$ is an equilibrium one \cite{noi}. 
This means that, given an equilibrium state, the nearest metastable state that
can be seen with this method is at
 overlap $\bar q$ with it and the nearest equilibrium state is at overlap 
$q^\star$.

\noindent
We can distinguish three different ranges:

{\parindent=1 truecm

\item{\bf i.} $q_{12}<q^\star$: from Figure 1 we can see that in this range there are only
two saddles, $S_1$ and $S_2$, and the path which links the two minima is then
$$
M_1\rightarrow S_1 \rightarrow S_2 \rightarrow M_2  \ .
\eq{buscadero}
$$
Since the highest  saddle is $S_1$, we have 
$$
\Delta F \geq \ V_3(S_1) - V_3(M_1)    \ .
\eq{nuto}
$$
It is worth to note that the value of $V_3$ in $M_1$ and
$S_1$ does not depend on $q_{12}$ and therefore in this range the barrier
is constant. In \cite{franzparisi} it has been introduced a two replica
potential $V_2$, function of the distance $q_{12}$ of replica 2  from the
equilibrium state of replica 1. $V_2$ has two minima corresponding to 
equilibrium states with zero mutual overlap: the most natural hypothesis is 
then that the maximum se\-pa\-ra\-ting these minima represents a first estimate of 
the free energy barrier between remote equilibrium states. It turns out that 
this estimate coincides with \re{nuto}, in agreement with the small value of
$q_{12}$ in this range. 
\item{\bf ii.} $q^\star < q_{12} < q'$: at $q^\star$ $S_2$ becomes higher than $S_1$
(see Figure 2) and so now the path is the same as in \re{buscadero}, 
but the barrier is
$$
\Delta F \geq \ V_3(S_2;q_{12}) - V_3(M_1)
\eq{oraziu}
$$
that depends on $q_{12}$. We note from Figure 2 that in this range there is a 
value of $q_{12}$ at which a new saddle $S$ appears together with a maximum;
however, since for $q_{12} < q'$ we have $V_3(S) > V_3(S_2)$, the path 
that minimizes the variation of $V_3$ still is \re{buscadero}.

\eject

\includegraphics{frango-fig1.ps}

\vbox{
      \hbox{     \vbox{\vglue 2.7 truecm}
                 \vbox{\vglue 7.0 truecm}                             }
      \hbox{     \hglue 5.5 truecm                                    }
      \vbox{\hsize=15 truecm \baselineskip=10 pt    
      \piccolo{ \noindent {\pbf Figure 1}: 
 		 The contour lines of $\s V_3$  for  $\s q_{12}=0.25$ and 
		 $\s \beta=1.64$.
                 The  axis are $\s x={1 \over \sqrt{2}}(q_{13}+q_{23})$
		 in the range $\s [0.25:0.6]$,
		 and $\s y={1 \over \sqrt{2}}(q_{13}-q_{23})$ in the range
                 $\s [-0.3,0.3]$. 
                 The two minima are visible on the right: 
                 $\s M_1$ (up) and $\s M_2$
                 (down); on the left the two saddles: $\s S_1$ (up) and $\s 
                 S_2$ (down). }                      }                    }
\vskip 0.2 truecm

\includegraphics{frango-fig2.ps}

\vbox{
      \hbox{     \vbox{\vglue 2.7 truecm}
                 \vbox{\vglue 7.0 truecm}                             }
      \hbox{     \hglue 5.5 truecm                                    }
      \vbox{ \hsize=15 truecm   \baselineskip=10 pt 

\piccolo{ \noindent   {\pbf Figure 2}: 
		  The contour lines of $\s V_3$  for  $\s q_{12}=0.36$.
		  The axis are the same as in Figure 1, with ranges 
                  $\s x\in [0.35,0.75]$ and 
                  $\s y\in [-0.2,0.2]$. The new saddle $\s S$ is clearly 
                  visible on the right, in the middle between the two 
                  minima.  }    }
                                                                           }
\vskip 0.2 truecm

\item{\bf iii.} 
$q'<q_{12}<\bar q$: at $q'$ the path that minimizes the variation
of $V_3$ suddenly changes because $S$ becomes lower than $S_2$; 
so we have
$$
M_1\rightarrow S \rightarrow M_2
\eq{dapaura}
$$
and 
$$
\Delta F \geq \ V_3(S;q_{12})-V_3(M_1)  \ .
\eq{trippa}
$$ 
This situation holds up to $q_{12}=\bar q$, when $M_2$ disappears merging
with $S$.
\par}
\vskip 0.3 truecm
The whole behaviour of $V_3$ in the saddles is shown in Figure 3.

\includegraphics{frango-fig3.ps}

\vbox{
      \hbox{ \hglue 1.3 truecm    \vbox{$\s V_3$ \vglue 2.7 truecm}
                 \vbox{\vglue 6.5 truecm}                             }
      \hbox{     \hglue 7.0 truecm $\s q_{12}$                        }
      \vbox{ \vglue 0.06 truecm}
      \vbox{ \hsize=15 truecm  \baselineskip=10 pt   

\piccolo{\noindent  {\pbf Figure 3}: 
                   $\s V_3$ evaluated in the three saddles $\s S_1$, $\s S_2$
                   and $\s S$ as a function of $\s q_{12}$. For 
                   $\s \beta=1.64$, $\s q^\star = 0.295$ and 
                   $\s q'=0.362$. }                    }
                                                                           }
\vskip 0.2 truecm

Finally, Figure 4 shows the behaviour of our estimate of the free energy 
barrier.   

\includegraphics{frango-fig4.ps}

\vbox{
      \hbox{ \hglue 1.2 truecm    \vbox{$\s \Delta F$ \vglue 2.7 truecm}
                 \vbox{\vglue 6.5 truecm}                             }
      \hbox{     \hglue 7.0 truecm $\s q_{12}$                                    }
      \vbox{ \hsize=15 truecm  \baselineskip=10 pt 

\piccolo{\noindent    {\pbf Figure 4}: 
		   The free energy barrier $\s \Delta F$ as a function of 
                   $\s q_{12}$.  }                    }
                                                                           }
\vskip 0.2 truecm

\autosez{conclu} Conclusions.
\par

Before analyzing these results it is useful to consider some properties of 
the potential in the stationary points. 

In the computation of $V_3$ we introduced  the overlap matrix  
$Q^{33}_{ab}$  of replica 3, and assumed for it a one step RSB form, 
with parameters $(x_s,s_1,s_0)$, where $x_s$ is the breaking point and
$s_1 \geq s_0$ are the overlaps. 

The physical meaning of breaking or not the replica symmetry is the usual 
one adapted to this context. An RS form of the overlap matrix can be 
associated to two very different situations: the first one corresponds to
a system which finds an exponentially high number of states (as in the 
$p$-spin model for $T_c<T<T_d$), while in the second one the phase space 
consists of just one state (as in the paramagnetic case). On the other
side, an RSB form means that the phase space is dominated by a number 
of order $N$ of states (as in the $p$-spin model for $T<T_c$).

It can be shown that in the minima of the potential it holds $s_1=s_0$, 
i.e. $Q^{33}$ turns out to be replica symmetric. This is what we expect: 
in $M_2$, for example, $q_{23}$ is big enough to constrain replica 3 to 
see just one state, the one of replica 2, and for this reason $Q^{33}$ is 
symmetric, $s_1=s_0$ representing the self overlap of the state. 

This symmetry is a very special feature of this kind of situation; indeed,
in the saddles $S_1$ and $S_2$  the matrix $Q^{33}$ is actually broken, 
meaning that the number of states accessible to replica 3 is in this 
case of order $N$ and that replica 3 does not thermalize into any of them.  

Surprisingly enough, in the saddle $S$ the matrix $Q^{33}$ is symmetric again;
moreover, in this saddle the free energy $f_3$ of replica 3 and the overlap 
$s_1=s_0$ satisfy  the relation between free energy and self overlap of TAP 
solutions and fulfil the stability condition with respect to
fluctuations of the overlap (longitudinal stability condition) of
\ccite{kpz}{crisa1}. These facts suggest that $S$ corresponds to a well 
defined TAP solution, but that, despite of the 
longitudinal stability condition, this solution does not identify a stable 
state, i.e. it is not a minimum of the TAP free energy. 
We can then make the hypothesis that $S$ represents a {\it real} saddle in the
phase space and that 
it is just the saddle which links the two states corresponding to $M_1$ and 
$M_2$; this conjecture is supported by the geometrical position of $S$ right 
between $M_1$ and $M_2$ (see Figure 2).
If this were true, in the range in which the path is $M_1 \rightarrow
S \rightarrow M_2$, expression \re{trippa} would give the exact free 
energy barrier, not only a lower bound.   

We note that the existence among TAP solutions of saddles which satisfy the
longitudinal stability condition of \ccite{kpz}{crisa1} requires a more 
complete analysis of the stability of TAP equations for this model. Moreover, 
it
would be interesting to make a computation of the complexity of these saddles,
to single out their contribution to the dynamics of the system.

\vskip 2 truecm
\noindent {\bf References.}
\vskip 0.5 truecm


\biblitem{ea} S.F. Edwards, P.W. Anderson, {\it J. Phys.} F {\bf 5} (1975), 
965.

\biblitem{sk} D. Sherrington, S. Kirkpatrick, {\it Phys. Rev. Lett.} {\bf 35}
(1975), 1792.

\biblitem{tap} D.J. Thouless, P.W. Anderson, R.G. Palmer, {\it Philos.
Mag.} {\bf 35} (1977), 593.

\biblitem{rsb1} G. Parisi, {\it Phys. Rev. Lett.} {\bf 23} (1979), 1754; 
                          
\biblitem{rsb2} G. Parisi, {\it J. Phys. A} {\bf 13} (1980), L115; 
                                           
\biblitem{rsb3} G. Parisi, {\it J. Phys. A} {\bf 13} (1980), 1887.

\biblitem{sompozip} H. Sompolinsky, A. Zippelius, {\it Phys. Rev.} B {\bf 25}
(1982), 6860.

\biblitem{ck1} L.F. Cugliandolo, J. Kurchan, {\it Phys. Rev. Lett.}
{\bf 71} (1993), 173.

\biblitem{tirumma} T.R. Kirkpatrick, D. Thirumalai, {\it Phis. Rev.} B {\bf 36} 
(1987), 5388.

\biblitem{crisa1} A. Crisanti, H.J. Sommers, {\it Z. Phys.} B {\bf 87} (1992), 
341.

\biblitem{crisa2} A. Crisanti, H. Horner, H.J. Sommers, {\it Z. Phys.} B
{\bf 92} (1993), 257.

\biblitem{ck2} L.F. Cugliandolo, J. Kurchan, {\it J. Phys.} A {\bf 27} 
(1994), 5749.

\biblitem{kpz} J. Kurchan, G. Parisi, M.A. Virasoro, {\it J. Phys. I France}
 {\bf 3} (1993), 1819.

\biblitem{franzparisi}  S. Franz, G. Parisi, {\it J. Phys. I France}
{\bf 5} (1995), 1401.

\biblitem{ferrero} M.E. Ferrero, M.A. Virasoro, {\it J. Phys. I France }
{\bf 4} (1994), 1819.

\biblitem{buribarrameza} A. Barrat, R. Burioni, M. M\'ezard, {\it J. Phys} A 
{\bf 29} (1996), L81.

\biblitem{monasson} R. Monasson, {\it Phys. Rev. Lett.}
{\bf 75} (1995), 2847.

\biblitem{crisatap} A. Crisanti, H.J. Sommers, {\it J. Phys. I France}
{\bf 5} (1995), 805.

\biblitem{mezpa} M. M\'ezard, G. Parisi, {\it J. Phys. A }
{\bf 23} (1990), L1229; {\it J. Phys. I France }
{\bf 1} (1991), 809.

\biblitem{nieu} Th.M. Nieuwenhuizen, {\it Phys. Rev. Lett. }
{\bf 74} (1996), 4289.

\biblitem{franz} S. Franz, private communication.

\biblitem{vira} M.A.Virasoro, {\sl Simulated annealing methods under analytical 
control}, in Pro\-cee\-dings of 19th Intl. Conf. on Stat. Phys. (IUPAP),
Xiamen, China; to be published.

\biblitem{grome} D.J. Gross, M. M\'ezard, {\it Nucl. Phys. B }
{\bf 240} (1984), 431.

\biblitem{gard} E. Gardner, {\it Nucl. Phys. B}
{\bf 257} (1985), 747.

\biblitem{noi} A. Cavagna, I. Giardina, G. Parisi, cond-mat 9611068,
submitted to {\it J. Phys. A}.

\insertbibliografia

\bye